\documentclass[twocolumn,superscriptaddress,preprintnumbers,amsmath,amssymb,prb]{revtex4}
\usepackage{graphicx}
\usepackage{dcolumn}
\usepackage{bm}
\usepackage[utf8]{inputenc}
\usepackage[T1]{fontenc}
\usepackage{mathptmx}
\usepackage{etoolbox}
\usepackage{amsmath}\usepackage{slashed}
\usepackage{mathtools}
\usepackage{color}
\usepackage{txfonts}
\UseRawInputEncoding

\begin{document}
\newcommand{\tm}{}

\title[Role of the Casimir force in the capacitive radio-frequency
microelectromechanical switches]{Role of the Casimir force in the capacitive radio-frequency
microelectromechanical switches}

\author{
G.~L.~Klimchitskaya{$^*$}}
\affiliation{Central Astronomical Observatory at Pulkovo of the Russian Academy of Sciences, St.Petersburg,
196140, Russia}
\affiliation{Peter the Great Saint Petersburg
Polytechnic University, Saint Petersburg, 195251, Russia}
\email{g.klimchitskaya@gmail.com}

\author{A.~S.~Korotkov}
\affiliation{Peter the Great Saint Petersburg
Polytechnic University, Saint Petersburg, 195251, Russia}

\author{V.~V.~Loboda}
\affiliation{Peter the Great Saint Petersburg
Polytechnic University, Saint Petersburg, 195251, Russia}

\author{
V.~M.~Mostepanenko}
\affiliation{Central Astronomical Observatory at Pulkovo of the Russian Academy of Sciences, St.Petersburg,
196140, Russia}
\affiliation{Peter the Great Saint Petersburg
Polytechnic University, Saint Petersburg, 195251, Russia}


\begin{abstract}
We determine the role of the fluctuation-induced Casimir force
acting between a membrane of cylindrical shape and a bottom
electrode in microelectromechanical capacitive switches. For
this purpose, the Casimir force is computed taking into account
the real properties of both a membrane and a bottom electrode
materials with account of surface roughness. The obtained
results are compared with those found for the smooth surfaces
using the idealization of ideal metal. It is shown that an
account of both the real material properties and surface
roughness is crucial for obtaining the correct values of
the Casimir force. According to our results, at the shortest
separations, when the switch membrane is in contact with the
transmission line, the magnitudes of the Casimir force may
exceed the magnitudes of the electric one
\tm{depending on the value of the operating voltage.}
The obtained values
of the Casimir force can be used for determining the thickness
of the switch membrane, which ensures the necessary magnitude
of the restoring elastic force required for a stable cyclic
functioning of the micromechanical switch with no pull-in.
\end{abstract}

\maketitle

\section{INTRODUCTION}

In the last few decades, the micromechanical systems (MEMS) play
a broad spectrum of roles in both a physical laboratory and in
nanotechnological applications, such as sensors, switches,
optical and cellular communications, etc.\cite{1} The MEMS
devices are typically operated by the combined action of the
electric and elastic forces. However, with reducing the
characteristic separations between the MEMS elements to below
a micrometer, the forces of quantum nature take effect. These
are the van der Waals and Casimir forces resulting from the
zero-point and thermal electromagnetic fluctuations.\cite{2,3}
The fluctuation-induced forces may play both the detrimental
role in functioning the MEMS devices by leading to their pull-in
instability and be advantageous by serving as one more operating
interaction (see Refs.~\onlinecite{3,4,5} for a review). A major
contribution to the investigation of the fluctuation-induced
forces in nanoscale was made by Prof. Dr. Rudolf Podgornik
(see, for instance, Refs.~\onlinecite{6,7,8,9}).

A highly important place among the MEMS devices is occupied by the
MEMS switches. The switches are \tm{always used} for connecting
and disconnecting electrical circuits. A variety of different MEMS
switches has been exploited. The most of them can be classified
under two types, resistive and capacitive.\cite{10,11,12} In the
resistive MEMS switches, an electrical circuit is closed in the
state of a metal-to-metal contact (low resistance) and open with
no contact (high resistance). In the capacitive MEMS switches, the
radio-frequency (RF) signal flows through the transmission line in
the state when a membrane is spaced above a central signal conductor
(low capacitance) and is terminated when a membrane contacts the
transmission line (high capacitance).

For the resistive MEMS switches, a consideration of the Casimir force
takes on great significance when the metal-to-metal gap decreases to
almost zero, so that the electrical circuit becomes closed. The role
played by the Casimir force in this situation \tm{in resistive switches}
was investigated in many
details.\cite{13,14,15,16,17,18,19,20,21,22,23,24,26,27,28,29,30,31,32,33,34}
The obtained results were used in a search for the stable operation
conditions of MEMS switches with no pull-in caused by the Casimir force.

Recent trends are toward increased use of the capacitive RF MEMS switches
which are highly reliable and admit operation up to a million cycles.
For the capacitive MEMS switches, the Casimir force becomes essential
when the switch membrane approaches the transmission line. However, in
spite of the fact that the capacitive switches and allied pull-in
phenomena are actively investigated in the last few
years,\cite{35,36,37,38,39,40,41,42,43,44,45,46,47,48,49,50,51,52,53,54,55,56,57,58,59,60,61,62}
no consideration has been given so far to the role of the Casimir force
in this kind of switches with account of a nonplanar shape of the membrane.

In this paper, we consider the capacitive RF MEMS switch incorporating the
rectangular membrane with fixed sides, which takes a cylindrical form
under an impact of the electric, elastic, and Casimir forces. This membrane
consists of two metallic layers and is situated above a plane conductor
covered with a dielectric plate. Using the Lifshitz theory, the
\tm{phenomenological}
method of additive summation of interatomic van der Waals
potentials corrected for the effects of
nonadditivity and the proximity force approximation, we calculate the
Casimir and electric forces in this configuration with account of surface
roughness. It is shown that at the
shortest membrane-plate separation, yielding the maximum capacitance and
termination of the RF signal flow, the Casimir force reaches the same
order of magnitude as the operating electric force.
\tm{Whether the magnitude of the Casimir force at the shortest membrane-plate separation is larger than the magnitude of electric one or not, the restoring elastic force should be larger than the magnitude of the Casimir force in order the membrane would be capable to return to its initial  position with no pull-in after the electric force is switched off.}
The condition on the
magnitude of restoring elastic force, required for avoiding the
pull-in and irreversible termination of the RF signal flow when the capacitive switch
loses its functionality, is obtained.

The paper is organized as follows. In Sec. II, the mathematical expressions
for the Casimir and van der Waals forces acting between a layered metallic
membrane of cylindrical shape and a bottom metallic electrode covered with
a dielectric plate are derived. In Sec. III, the expression for the electric
force in the same configuration is obtained. Section IV is devoted to the
impact of surface roughness on the Casimir, van der Waals and electric
forces. In Sec. V, the Casimir, van der Waals and electric forces acting in
the capacitive RF MEMS switches with a cylindrical membrane are computed for
the realistic values of switch parameters. Section VI contains our
conclusions and a discussion of the obtained results.

\section{Casimir force between cylindrical membrane of a capacitive switch
and bottom electrode}
\newcommand{\kb}{{k_{\bot}}}
\newcommand{\xk}{{(i\xi_l,k_{\bot}\!)}}
\newcommand{\ve}{\varepsilon}

\begin{figure}[b]
\vspace*{-2.5cm}
\hspace*{-2.7cm}
\includegraphics[width=5.0in]{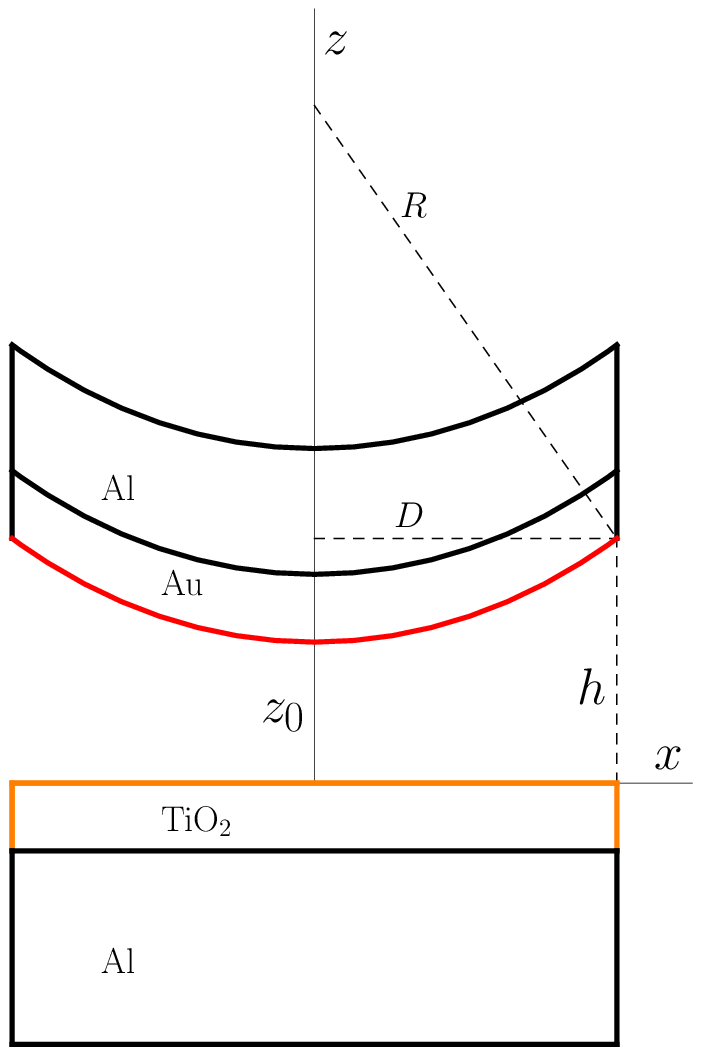}
\vspace*{-5.4cm}
\caption{\label{fg1}
 Schematic of the geometrical configuration of MEMS capacitive switch
with a cylindrical membrane shown not to scale (see the text for further
discussion).
}
\end{figure}
The geometry of a capacitive switch is schematically shown in Fig.~\ref{fg1} as a cross
section by the plane $y=0$ (not to scale).  The top cantilever membrane  is of rectangular
shape with the width $2D$ and length $L$ extended along the $y$ axis. It is a metallic plate
with fixed sides made of Al coated from the underside with the thin Au protective layer.
In the absence of external forces, the switch membrane is a flat plate situated in the plane
$z={\rm const}$. The bottom electrode, which serves as the central signal line, is also made of Al
and covered by a dielectric plate in order to increase capacitance made, for instance, of
TiO$_2$. The thicknesses of Au layer and TiO$_2$ plate are $d^{(+1)}$ and $d^{(-1)}$,
respectively. The thicknesses of Al plates are much larger than 100~nm, so that they can be
considered as semispaces when calculating the Casimir force.\cite{3}
The dielectric (usually SiO$_2$) substrate underlying the Al plate does not influence the
Casimir force and is not shown in Fig.~\ref{fg1}. Under the impact of dc voltage applied
between the membrane and bottom electrode, the membrane surface assumes the cylindrical
form.\cite{39,40,43,49,53,59,62a,63,64}

\subsection{Fluctuation-induced forces at sufficiently large separations}

According to the proximity force approximation (PFA), under the condition $z_0\ll R$,
where $z_0$ is the closest separation between a membrane and a bottom electrode and
$R=R(z_0)$ is the membrane curvature radius, the membrane can be replaced with a set
of plane elementary plates parallel to the plane $xy$. In doing so, the Casimir force
between the membrane and the bottom electrode can be approximately represented as
\begin{equation}
F_C(z_0)=\int_{\sigma}P\left(z(x)\right)\,dxdy,
\label{eq1}
\end{equation}
\noindent
where $P\left(z(x)\right)$ is the Casimir pressure between each elementary plane plate
and the underlying element of the bottom electrode,
$z(x)=R(z_0)+z_0-[R^2(z_0)-x^2]^{1/2}$ is the equation of lower boundary surface of
the membrane, and $\sigma$ is the projection of this surface to the plane $z=0$.
Note that the magnitude of corrections to Eq.~(\ref{eq1}) is much below the ratio
$z_0/R(z_0)$,\cite{65,66,67} which is much less than unity (see Sec.~V for the specific
values of all geometrical parameters characteristic for the capacitive MEMS switches).

The Casimir pressure $P\left(z(x)\right)$ between the pairs of plane parallel plates
spaced $z(x)$ apart can be expressed by means of the Lifshitz formula\cite{3,68}
\begin{eqnarray}
&&
P\left(z(x)\right)=-\frac{k_BT}{\pi}\sum_{l=0}^{\infty}
{\vphantom{\sum}}^{\prime}\int_0^{\infty}q_l\kb d\kb
\nonumber\\
&&~~
\times\sum_{\alpha}\frac{R_{\alpha}^{(+)}\xk R_{\alpha}^{(-)}\xk
e^{-2q_lz(x)}}{1-R_{\alpha}^{(+)}\xk R_{\alpha}^{(-)}\xk
e^{-2q_lz(x)}},
\label{eq2}
\end{eqnarray}
\noindent
where $k_B$ is the Boltzmann constant, $T$ is the temperature,
$q_l=(k_{\bot}^2+\xi_l^2/c^2)^{1/2}$, $\kb$ is the projection of the wave vector
on the plane $z=0$, $\xi_l=4\pi^2 k_BTl/h$, $l=0,\,1,\,2,\,\ldots,$ are the Matsubara
frequencies, the prime on the summation sign divides the term with $l=0$ by 2, and
$R_{\alpha}^{(\pm)}\xk$ are the reflection coefficients on the boundary surfaces
of the membrane and the bottom electrode, respectively, for the transverse magnetic
($\alpha={\rm TM}$) and transverse electric ($\alpha={\rm TE}$) polarizations of
the electromagnetic field. Note that the Casimir force between a one layer cylinder
with fixed $R$ and a plate using the PFA method was considered in Ref.~\onlinecite{69}.

The reflection coefficients on the layered structures of a membrane and an electrode
are expressed via the relative dielectric permittivities of layer materials.
We notate the permittivities of membrane material (Al and Au) $\ve^{(+1)}(\omega)$ and
 $\ve^{(+2)}(\omega)$, respectively, the permittivities of the bottom electrode
 (TiO$_2$ and Al)  $\ve^{(-1)}(\omega)$ and  $\ve^{(-2)}(\omega)$, respectively,
 and $\ve^{(0)}=1$ the permittivity of the gap material (air). In this specific case
  $\ve^{(+2)}(\omega)=\ve^{(-2)}(\omega)$. Then the reflection coefficients
$R_{\rm TM,TE}^{(\pm)}$ are expressed as\cite{3}
\begin{equation}
R_{\rm TM,TE}^{(\pm)}\xk=\frac{r_{\rm TM,TE}^{(0,\pm 1)}\xk+
r_{\rm TM,TE}^{(\pm 1,\pm 2)}\xk e^{-2k_l^{(\pm 1)}d^{(\pm 1)}}}{1+
r_{\rm TM,TE}^{(0,\pm 1)}\xk
r_{\rm TM,TE}^{(\pm 1,\pm 2)}\xk e^{-2k_l^{(\pm 1)}d^{(\pm 1)}}},
\label{eq3}
\end{equation}
\noindent
where
\begin{equation}
r_{\rm TM}^{(n,n^{\prime})}\xk=\frac{\ve_l^{(n^{\prime})}k_l^{(n)}-
\ve_l^{(n)}k_l^{(n^{\prime})}}{\ve_l^{(n^{\prime})}k_l^{(n)}+
\ve_l^{(n)}k_l^{(n^{\prime})}},
\nonumber
\end{equation}
\begin{equation}
r_{\rm TE}^{(n,n^{\prime})}\xk=\frac{k_l^{(n)}-k_l^{(n^{\prime})}}{k_l^{(n)}+
k_l^{(n^{\prime})}},
\label{eq4}
\end{equation}
\noindent
$n=0,\,\pm1,\,\pm2$, and
\begin{equation}
k_l^{(n)}=k_l^{(n)}\xk=\left[k_{\bot}^2+\ve_l^{(n)}\frac{\xi_l^2}{c^2}
\right]^{1/2}\!\!\!\!\!\!\!, \quad \ve_l^{(n)}=\ve^{(n)}(i\xi_l).
\label{eq5}
\end{equation}

Now we expand the fraction under the sum in $\alpha$ in Eq.~(\ref{eq2}) in powers
of $R_{\alpha}^{(+)}R_{\alpha}^{(-)}\exp(-2q_lz)$ and substitute Eq.~(\ref{eq2})
in Eq.~(\ref{eq1})
\begin{widetext}
\begin{eqnarray}
&&
F_C(z_0)=-\frac{2k_BTL}{\pi}\sum_{l=0}^{\infty}
{\vphantom{\sum}}^{\prime}\int_0^{\infty}q_l\kb d\kb
\nonumber\\
&&~~\times\sum_{m=1}^{\infty}\left[{R_{\rm TM}^{(+)}}^m\xk
{R_{\rm TM}^{(-)}}^m\xk+{R_{\rm TE}^{(+)}}^m\xk
{R_{\rm TE}^{(-)}}^m\xk\right]
\,\int_{0}^{D}dxe^{-2mq_lz(x)}.
\label{eq6}
\end{eqnarray}
\end{widetext}

First we calculate the integral
\begin{equation}
I\equiv\int_0^D dx e^{-2mq_lz(x)},
\label{eq7}
\end{equation}
\noindent
entering Eq.~(\ref{eq6}). This can be done under the conditions
\begin{equation}
h\ll D\ll R(z_0),
\label{eq8}
\end{equation}
\noindent
which is satisfied for the capacitive switches under consideration, taking into
account that the Casimir forces have a profound impact on the functioning of MEMS
switches under a condition $z_0\ll h$ (see Sec.~V).

{}From Fig.~\ref{fg1}, it is seen that
\begin{equation}
R^2(z_0)=D^2+[R(z_0)-(h-z_0)]^2,
\label{eq9}
\end{equation}
\noindent
i.e.,
\begin{equation}
2R(z_0)(h-z_0)=D^2+(h-z_0)^2\approx D^2.
\label{eq10}
\end{equation}
\noindent
{}From this it follows that
\begin{equation}
R(z_0)\approx\frac{D^2}{2(h-z_0)}.
\label{eq11}
\end{equation}

Using the equation of the membrane bottom boundary surface [see below Eq.~(\ref{eq1})]
and taking into account that $x\leqslant D$, with the help of Eq.~(\ref{eq8}) one finds
\begin{equation}
z(x)\approx R(z_0)+z_0-R(z_0)\left[1-\frac{x^2}{2R^2(z_0)}\right]=
z_0+\frac{x^2}{2R(z_0)}.
\label{eq12}
\end{equation}
\noindent
Substituting this in Eq.~(\ref{eq7}), one arrives at
\begin{equation}
I=\exp(-2mq_lz_0)\int_0^D\!dx\,\exp\left[-mq_l\frac{x^2}{R(z_0)}\right],
\label{eq13}
\end{equation}
\noindent
and, introducing the new integration variables in Eqs.~(\ref{eq6}) and (\ref{eq13}),
$v=2q_lz_0$ and $u=x/R(z_0)$, at\cite{70}
\begin{eqnarray}
&&
I=r(z_0)\exp(-mv)\int_0^{D/R(z_0)}\!du\,\exp\left[-m\frac{R(z_0)}{2z_0}vu^2\right]
\nonumber \\
&&~~
=R(z_0)e^{-mv}\sqrt{\frac{\pi z_0}{2mR(z_0)v}}\,{\rm erf}\left(D
\sqrt{\frac{mv}{2z_0R(z_0)}}\right),
\label{eq14}
\end{eqnarray}
\noindent
where ${\rm erf}(z)$ is the Gauss error function.

Rewriting Eq.~(\ref{eq6}) in terms of the integration variable $v$ and substituting
there the value of $I$ from Eq.~(\ref{eq14}), we obtain
\begin{eqnarray}
&&
F_C(z_0)=-\frac{k_BTL}{(2z_0)^{5/2}}\sqrt{\frac{R(z_0)}{\pi}}
\sum_{l=0}^{\infty}{\vphantom{\sum}}^{\prime}\int_{\zeta_l}^{\infty}
v^{3/2}\,dv\,\sum_{m=1}^{\infty}\frac{e^{-mv}}{\sqrt{m}}
\nonumber\\
&&~~\times
\left[{R_{\rm TM}^{(+)}}^m(i\zeta_l,v) {R_{\rm TM}^{(-)}}^m(i\zeta_l,v)
+{R_{\rm TE}^{(+)}}^m(i\zeta_l,v) {R_{\rm TE}^{(-)}}^m(i\zeta_l,v)\right]
\nonumber\\
&&~~\times
{\rm erf}\left(D\sqrt{\frac{mv}{2z_0R(z_0)}}\right),
\label{eq15}
\end{eqnarray}
\noindent
where $\zeta_l=\xi_l/\omega_c=2z_0\xi_l/c$, the reflection coefficients
$R_{\rm TM,TE}^{(\pm)}(i\zeta_l,v)$ are expressed via the quantities
$r_{\rm TM,TE}^{(n,n^{\prime})}(i\zeta_l,v)$ by Eq.~(\ref{eq3}) and these
quantities are expressed via $k^{(n)}(i\zeta_l,v)$ by Eq.~(\ref{eq4}).
In so doing $\xi_l$ and $\kb$ are replaced with $\zeta_l$ and $v$,
respectively, whereas $k^{(n)}(i\zeta_l,v)$ are given by
\begin{equation}
k_l^{(n)}=k^{(n)}(i\zeta_l,v)=\frac{1}{2z_0}
\left[v^2+(\ve_l^{(n)}-1)\zeta_l^2\right]^{1/2}
\label{eq16}
\end{equation}
\noindent
with $\ve_l^{(n)}=\ve^{(n)}(i\omega_c\zeta_l)$.

Now we note that, due to the exponential factor, the main contribution to the integral
in Eq.~(\ref{eq15}) is given by the domain $mv\sim 1$. Using this fact and Eq.~(\ref{eq11})
under a condition $z_0\ll h$, the argument of the error function in Eq.~(\ref{eq15}) is
estimated as
\begin{equation}
z\equiv D\sqrt{\frac{mv}{2z_0R(z_0)}}=\sqrt{\frac{mvD^2}{2z_0R(z_0)}}
\approx\sqrt{mv\frac{h}{z_0}}\gg 1.
\label{eq17}
\end{equation}

Taking this into account, one can use the asymptotic expression of large argument \cite{70}
\begin{equation}
\lim_{z\to\infty}{\rm erf}(z)=1
\label{eq18}
\end{equation}
\noindent
in Eq.~(\ref{eq15}) and, finally, rewrite this equation in terms of the polylogarithm
functions \cite{71}
\begin{eqnarray}
&&
F_C(z_0)=-\frac{k_BTL}{(2z_0)^{5/2}}\sqrt{\frac{R(z_0)}{\pi}}
\sum_{l=0}^{\infty}{\vphantom{\sum}}^{\prime}\int_{\zeta_l}^{\infty}
v^{3/2}\,dv\,
\nonumber\\
&&~~~~~\times\left\{{\rm Li}_{1/2}
\left[{R_{\rm TM}^{(+)}(i\zeta_l,v)}{R_{\rm TM}^{(-)}(i\zeta_l,v)}\,e^{-v}\right]
\right.
\nonumber\\
&&~~~~~~~~~~~\left. +
{\rm Li}_{1/2}\left[{R_{\rm TE}^{(+)}(i\zeta_l,v)} {R_{\rm TE}^{(-)}(i\zeta_l,v)}\,
e^{-v}\right]\right\}.
\label{eq19}
\end{eqnarray}

In Sec.~V, this equation is used for computations of the Casimir force acting in the
capacitive MEMS switch on its membrane in close proximity to the bottom electrode.

\subsection{Fluctuation-induced force at shortest separations between
interacting surfaces}

The Lifshitz formula (\ref{eq2}) employed in the derivation of Eq.~(\ref{eq19}) describes the interacting surfaces by means of idealization of continuous dielectric permittivities.
This description is applicable only on separation distances between the surfaces much larger
that their lattice constants, i.e., at separations above $z_0\simeq 3~$nm
(see Ref.~\onlinecite{68} and a more
modern discussion on this point\cite{72,73}).

For the elucidation of the role of Casimir forces in capacitive MEMS switches, it is
desirable to calculate it at separations down to $z_0=1~$nm. This can be done using
the methods of molecular dynamics,\cite{74} but here we use a more simple
\tm{approximate }
phenomenological
method of the pairwise summation of interatomic van der Waals potentials corrected for
accounting the effects of nonadditivity,
\cite{33,34,75,75a}
\tm{which provides a reasonable assessment of the force magnitude in the separation
region from 1 to 3~nm.}
The nonretarded van der Waals
interaction potential between a membrane atom and a bottom electrode atom spaced $r$
apart has the form\cite{2,3}
\begin{equation}
V_{\rm vdW}^{aa}(r)=-\frac{C}{r^6},
\label{eq20}
\end{equation}
\noindent
where the constant $C$ depends on the type of atoms.

First, we integrate the potential (\ref{eq20}) over the volume of the bottom electrode.
In doing so, due to the quick decrease of the potential (\ref{eq20}) with $r$,
only the TiO$_2$ dielectric plate of thickness $d^{(-1)}$ determines the total
result. Thus, this plate can be considered as infinitely thick. The resulting
atom-plate interaction potential and force are\cite{3,33}
\begin{equation}
V_{\rm vdW}^{ap}(z)=-\frac{\pi Cn_2}{6z^3},\qquad
F_{\rm vdW}^{ap}(z)=-\frac{\pi Cn_2}{2z^4},
\label{eq21}
\end{equation}
\noindent
where $z$ is the height of the membrane atom above a TiO$_2$ plate and $n_2$ is
the number of plate atoms per unit volume.

Now it is necessary to integrate Eq.~(\ref{eq21}) over the volume of a cylindrical
membrane. In this case, again, only the Au layer determines the total result and the
integration with respect to $z$ can be performed to infinity. The van der Waals
force acting on an elementary volume of Au layer above the area element $dS=Ldx$ is
\begin{eqnarray}
&&
dF_{\rm vdW}(z)=n_1\int_{z(x)}^{\infty}F_{\rm vdW}^{ap}{(z_1)}dz_1dS
\nonumber \\
&&~~~
=-\frac{1}{2}\pi n_1n_2CLdx\int_{z(x)}^{\infty}\frac{dz_1}{z_1^4} =
-\frac{\pi n_1n_2CLdx}{6z^3(x)},
\label{eq22}
\end{eqnarray}
\noindent
where $n_1$ in the number of Au atoms per unit volume of the membrane.

The total van der Waals force acting on the membrane is obtained by twice the integration of
Eq.~(\ref{eq22}) with respect to $x$ using Eq.~(\ref{eq12}) and Ref.~\onlinecite{70}
\begin{eqnarray}
&&
F_{\rm vdW}(z_0)=-\frac{1}{3}\pi n_1n_2CL\int_0^D
\frac{dx}{\left[R(z_0)+z_0-\sqrt{R^2(z_0)-x^2}\right]^3}
\nonumber \\
&&~~~\approx
-\frac{8}{3}\pi n_1n_2CLR^3(z_0)\int_0^{\infty}
\frac{dx}{\left[2z_0R(z_0)+x^2\right]^3}
\nonumber \\
&&~~~~~
=-\frac{1}{8\sqrt{2}}\pi^2n_1n_2CL\frac{1}{z_0^2}\sqrt{\frac{R(z_0)}{z_0}}.
\label{eq23}
\end{eqnarray}
\noindent
Note that in this calculation the upper integration limit $D$ was replaced with
infinity because the value of integral is determined by $x\ll D$.
\tm{It is significant that the dependence on separation in Eq. (23) obtained by
a summation of the interatomic van der Waals potentials is the same as was obtained
from the Lifshitz formula in Eq. (19).}

It is well known that the additive summation of interatomic forces overestimates the
van der Waals force acting between macrobodies.\cite{3,75,76} To approximately correct
the result (\ref{eq23}), it was suggested to divide it by the factor
\begin{equation}
K=\frac{F_{\rm vdW}(z_0=3\,\mbox{nm})}{F_{C}(z_0=3\,\mbox{nm})},
\label{eq24}
\end{equation}
\noindent
where  $F_{C}(z_0=3\,\mbox{nm})$ is calculated by Eq.~(\ref{eq19}). As a result,
the corrected value of the van der Waals force is given by
\begin{eqnarray}
&&
F_{\rm vdW}^{\rm corr}(z_0)=\frac{F_{\rm vdW}(z_0)}{K}=
\left(\frac{3\,\mbox{nm}}{z_0}\right)^{5/2}
\sqrt{\frac{R(z_0)}{R(z_0=3\,\mbox{nm})}}\,F_C(z_0=3\,\mbox{nm})
\nonumber \\
&&~~~~~\approx
\left(\frac{3\,\mbox{nm}}{z_0}\right)^{5/2}
\!\!\!F_C(z_0=3\,\mbox{nm}),
\label{eq25}
\end{eqnarray}
\noindent
taking into account that at $z_0\leqslant 3~$nm the sphere radius is practically constant.
At $z_0=3~$nm it holds $F_{\rm vdW}^{\rm corr}(z_0)=F_C(z_0)$.
\tm{In the corrected additive summation method, the material dependence of interacting
bodies is taken into account by means of the Casimir force value at 3~nm expressed
using the Lifshitz theory by Eq.~(19). It was shown\cite{75} that the theoretical results found
by this method are in rather good agreement with the measurement data down to separation
of 0.5~nm, i.e., to even shorter separations than those considered in this paper.}
\tm{In fact the corrected method of additive summation of the interatomic potentials
is of the same accuracy as the commonly used expression for the van der Waals force
in terms of the Hamaker constant, which independence on separation was confirmed
experimentally at least at separations down to 1.5~nm.\cite{75b,75c}}

In Sec.~V, the additional force of quantum nature induced by the electromagnetic
fluctuations, which acts in a MEMS capacitive switch when its membrane approaches
the bottom electrode, is computed using Eq.~(\ref{eq19}) at $z_0\geqslant 3~$nm
and Eq.~(\ref{eq25}) at $z_0<3~$nm.

\section{Electric force between cylindrical membrane of a capacitive switch
and bottom electrode}

\newcommand{\eo}{{\epsilon_{\,0}}}

The electric force acting in the MEMS capacitive switch with a cylindrical membrane
can also be calculated using the PFA. For a plane membrane separated by an air gap of
thickness $h$ from the dielectric plate of thickness $d^{(-1)}$ the capacitance is
given by
\begin{equation}
C=\frac{\eo S}{h+\frac{d^{(-1)}}{\ve_0^{(-1)}}},
\label{eq26}
\end{equation}
\noindent
where $S=2DL$ is the area of the plates, $\eo$ is the permittivity of a free space,
and $\ve_0^{(-1)}=\ve^{(-1)}(0)$ is the relative static permittivity of the dielectric plate.

Then for a force acting in our plane capacitor one obtains
\begin{equation}
F_{\rm el}=-\frac{QE}{2}=-\frac{QU}{2(h+d^{(-1)})}=
-\frac{1}{2}\frac{U^2}{h+d^{(-1)}}
\frac{\epsilon_0S}{h+\frac{d^{(-1)}}{\ve_0^{(-1)}}},
\label{eq27}
\end{equation}
\noindent
where $U$ is the voltage applied between the membrane and bottom electrode.

Now we consider the point $z$ of the cylindrical membrane surface, the plane plate
of the area $Ldx$ and, by applying the PFA, obtain an expression for the electric force
acting between the cylindrical membrane and bottom electrode with the help of
Eq.~(\ref{eq27})
\begin{widetext}
\begin{equation}
F_{\rm el}(z_0)=-\eo LU^2\int_0^D
\frac{dx}{\left[R(z_0)+z_0-\sqrt{R^2(z_0)-x^2}+d^{(-1)}\right]\,
\left[R(z_0)+z_0-\sqrt{R^2(z_0)-x^2}+\tilde{d}^{(-1)}\right]},
\label{eq28}
\end{equation}
\end{widetext}
\noindent
where $\tilde{d}^{(-1)}=d^{(-1)}
\ve_0^{(-1)}$.

By introducing the dimensionless quantities
\begin{widetext}
\begin{equation}
\rho=\frac{x}{D},\qquad a=\frac{R(z_0)}{D}, \qquad
b_1=\frac{R(z_0)+z_0+d^{(-1)}}{D},
\qquad
b_2=\frac{R(z_0)+z_0+\tilde{d}^{(-1)}}{D},
\label{eq29}
\end{equation}
\noindent
Eq.~(\ref{eq28}) can be rewritten as
\begin{eqnarray}
&&
F_{\rm el}(z_0)=-\frac{\eo LU^2}{D}\int_0^!
\frac{d\rho}{(b_1-\sqrt{a^2-\rho^2})(b_2-\sqrt{a^2-\rho^2})}
\nonumber \\
&&~=
-\frac{2\eo LU^2}{D(b_2-b_1)}\left(\frac{b_1}{\sqrt{b_1^2-a^2}}
\arctan\frac{a-\sqrt{a^2-1}}{b_1-a}\right.
\left.-
\frac{b_2}{\sqrt{b_2^2-a^2}}
\arctan\frac{a-\sqrt{a^2-1}}{b_2-a}\right).
\label{eq30}
\end{eqnarray}

After returning to the original dimensional variables, the electric  force takes
the following final form
\begin{eqnarray}
&&
F_{\rm el}(z_0)=-\frac{2\eo \ve_0^{(-1)}LU^2}{d^{(-1)}(\ve_0^{(-1)}-1)}
\left\{
\frac{R(z_0)+z_0+\tilde{d}^{(-1)}}{\sqrt{(z_0+\tilde{d}^{(-1)})
[2R(z_0)+z_0+\tilde{d}^{(-1)}]}}\,
\arctan\frac{R(z_0)-\sqrt{R^2(z_0)-D^2}}{z_0+\tilde{d}^{(-1)}}\right.
\nonumber \\
&&~~~~~~~\left. -
\frac{R(z_0)+z_0+{d}^{(-1)}}{\sqrt{(z_0+{d}^{(-1)})
[2R(z_0)+z_0+{d}^{(-1)}]}}\,
\arctan\frac{R(z_0)-\sqrt{R^2(z_0)-D^2}}{z_0+{d}^{(-1)}}\right\}.
\label{eq31}
\end{eqnarray}
\end{widetext}

In Sec.~V, the magnitudes of electric force in the capacitive switch computed using
Eq.~(\ref{eq31}) are compared with the magnitudes of the Casimir force.

\section{Impact of surface roughness on the Casimir and electric forces}

The outlined theory of the Casimir and electric forces acting in the configuration
of MEMS capacitive switches relates to the perfectly smooth surfaces of a membrane
and a bottom electrode. In real situations, however, the interacting surfaces are
rough. This makes an impact on both the Casimir and electric forces and should be
taken into account in their calculation. Specifically, the case of electric force
acting between the rough surfaces in capacitive switches was already considered
in the literature.\cite{47,77,78}

Extensive studies of the role of surface roughness in measuring the Casimir force
were performed by many authors (see Refs.~\onlinecite{3,79,80} for a review).
It was shown that if the correlation length of stochastic roughness exceeds the
separation distance between the rough surfaces (this condition is satisfied at
short separations), the effect of roughness can be reliably taken into account
by using the PFA.

In doing so, the rough surfaces are characterized by some smooth mean levels,
so that the deviations in both sides are counted from these levels.
We notate these deviations from the mean roughness levels on a membrane and
a bottom electrode $H_i^{(+1)}$ and  $H_i^{(-1)}$, respectively.
Each deviation occurs with the probability $w_i^{(+1)}$ and  $w_i^{(-1)}$,
respectively,so that it holds
\begin{equation}
\sum_{i=1}^{N^{(n)}}w_i^{(n)}=1,\qquad
\sum_{i=1}^{N^{(n)}}H_i^{(n)}w_i^{(n)}=0,
\label{eq32}
\end{equation}
\noindent
where $n=\pm 1$, and $N^{(n)}$ are the chosen numbers of deviations from the
mean levels on each of the two surfaces. Within this approach, the surface
roughness is characterized by the amplitude and the root mean square deviation
from the mean roughness level which are defined as
\begin{equation}
A^{(n)}=\max|H_i^{(n)}|,\qquad
\delta_{\rm rms}^{(n)}=\sqrt{\sum_{i=1}^{N^{(n)}}|H_i^{(n)}|^2w_i^{(n)}},
\label{eq33a}
\end{equation}
\noindent
respectively.

Then, the Casimir and electric forces between the rough surfaces can be approximately
calculated as
\begin{equation}
F_{C,\,{\rm el}}^{R}(z_0)=\sum_{i=1}^{N^{(+1)}}\sum_{j=1}^{N^{(-1)}}
w_i^{(+1)}w_J^{(-1)}
F_{C,\,{\rm el}}(z_{ij}),
\label{eq33}
\end{equation}
\noindent
where $z_{ij}$ are the separation distances between the points of rough surfaces
in different locations.

To evaluate the role of surface roughness in calculation of the Casimir and electric
forces in the capacitive MEMS switches, we consider three models of stochastic roughness
on the Au surface of a membrane choosing the typical values of $H_i^{(+1)}$ and
$w_i^{(+1)}$ as were determined by means of an atomic force microscope in precision
experiments on measuring the Casimir force.\cite{3,79}
In the model 1, we put $H_1^{(+1)}=1~$nm and $H_2^{(+1)}=-2~$nm with the
probabilities  $w_1^{(+1)}=2/3$ and $w_2^{(+1)}=1/3$, respectively
(note that for each rough surface the positive value of the deviation $H_i$ is chosen in
the direction to the opposite rough surface).
Here $N^{(+1)}=2$.
In the model 2, $H_i^{(+1)}$ ($i=1,\,2,\,3,\,4$) take the values 2, 1, --1, --2~nm with
the respective  probabilities  $w_i^{(+1)}=1/6,\,1/3,\,1/3,\,1/6$ and $N^{(+1)}=4$.
In the model 3, $H^{(+1)}$ is equal to 2 and --1~nm, with the probabilities 1/3
and 2/3, respectively ($N^{(+1)}=2$).
For all the three models listed above, $A^{(+1)}=2~$nm and
$\delta_{\rm rms}^{(+1)}=1.4~$nm.

The surfaces of dielectric plates are usually more smooth than of metallic ones.
For instance, Si surfaces used in measurements of the Casimir force are
characterized by  $\delta_{\rm rms}=0.1~$nm.
For the model of roughness on the TiO$_2$ plate, we choose not so stringent requirement
and put $H_i^{(-1)}=\pm 1~$nm with equal probabilities  $w_i^{(-1)}=1/2$.
These results in $A^{(-1)}=1~$nm, $\delta_{\rm rms}^{(-1)}=1.4~$nm and $N^{(-1)}=2$.

By notating $z_0$ the closest separation between two rough surfaces, the separation
between any two levels of roughness is given by
\begin{equation}
z_{ij}=z_0+|H_i^{(+1)}-H_1^{(+1)}|+|H_j^{(-1)}-H_1^{(-1)}|).
\label{eq34}
\end{equation}
\noindent
Substituting this result in Eq.~(\ref{eq33}), for the Casimir and electric forces
acting between the rough surfaces of a membrane and a bottom electrode one
finally obtains
\begin{eqnarray}
&&
F_{C,\,{\rm el}}^{R}(z_0)=\sum_{i=1}^{N^{(+1)}}\sum_{j=1}^{N^{(-1)}}
w_i^{(+1)}w_j^{(-1)}
\nonumber \\
&&~~~~~\times
F_{C,\,{\rm el}}(z_0+|H_i^{(+1)}-H_1^{(+1)}|+|H_j^{(-1)}-H_1^{(-1)}|),
\label{eq35}
\end{eqnarray}
\noindent
where the Casimir force $F_C$ is given by Eqs.~(\ref{eq19}) and (\ref{eq25}),
whereas the electric force is expressed by Eq.~(\ref{eq31}).

\section{CALCULATION OF THE CASIMIR AND ELECTRIC FORCES IN CAPACITIVE
RF MEMS SWITCHES}

Now we compute both the Casimir (van der Waals) and electric forces acting
between the membrane and the bottom electrode of a capacitive switch.
These computations are
performed for the following typical parameters of the MEMS capacitive
switch: $L = 20~\muup$m, $D = 40~\muup$m, the thickness of Al membrane
is $1~\muup$m and of the protective Au layer on its bottom surface is
$d^{(+1)} = 10$~nm, the thickness of TiO$_2$ plate is $d^{(-1)} = 100$~nm
and of the bottom electrode made of Al is 500~nm.
The width of the gap between a membrane and a bottom electrode is
$h=2~\muup$m and, as it follows from Eq.~(\ref{eq11}), the minimum value
of the curvature radius of a membrane is $\min R=R(0)=400~\muup$m.
We recall that Al
surfaces can be considered as infinitely thick.

As indicated in Sec. II, to calculate the Casimir force, one needs to know
the values of the dielectric permittivities of Al, Au, and TiO$_2$ at the
pure imaginary Matsubara frequencies $i\xi_l$. For Al and Au these values
are found using the tabulated optical data for the complex indices of
refraction\cite{81} extrapolated to lower frequencies and the
Kramers-Kronig relation expressing $\varepsilon^{(n)}(i\xi_l)$ via
${\rm Im}\varepsilon^{(n)}(\omega)$ (see Refs.~\onlinecite{3,79} for
details).

\begin{figure}[b]
\vspace*{-4.cm}
\hspace*{-1.9cm}
\includegraphics[width=5.5in]{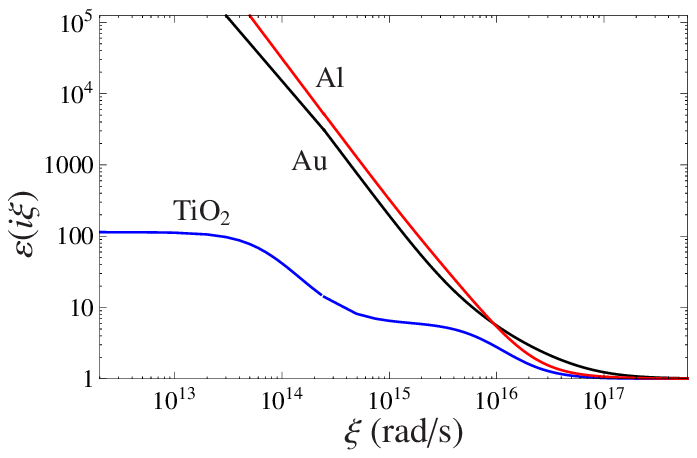}
\vspace*{-12.cm}
\caption{\label{fg2}
 The dielectric permittivities of Au, Al, and TiO$_2$ are shown as the
functions of frequency along the imaginary frequency axis.
}
\end{figure}

As was shown by numerous experiments performed by different experimental
groups (see the reviews\cite{3,79} and
the more recent experiments\cite{81a,82,83,84,85}), the Casimir forces computed
in this way are excluded by the measurement data if the
tabulated optical data are
extrapolated by means of the Drude model commonly used at low frequencies.
However, if the extrapolation is made using the plasma model, the obtained
theoretical Casimir forces are in good agreement with the measurement data.
Recently it was found both theoretically\cite{86} and experimentally\cite{87}
that the reason for this paradoxical situation is that the Drude model fails
to correctly describe the response of metals to the electromagnetic
field in the region of transverse electric evanescent waves contributing
just to the Casimir force. Therefore, here we use the experimentally
consistent extrapolation to low frequencies by means of the plasma model.
Note, however, that at separations $z_0 < 100$~nm, which are of our major
interest, the obtained values of the Casimir force do not depend on the
type of extrapolation used. In Fig.~\ref{fg2}, the obtained results for
$\varepsilon^{(1)}$ (Au) and $\varepsilon^{(2)} = \varepsilon^{(-2)}$ (Al)
are shown as the functions of frequency along the pure imaginary frequency
axis (the lines labeled Au and Al).

For TiO$_2$, which is the material of the dielectric plate in Fig.~\ref{fg1},
there is an accurate analytic representation for its dielectric permittivity
along the imaginary frequency axis\cite{88}
\begin{equation}
\ve^{(-1)}(i\xi)=1+\frac{\omega_{\rm IR}^2C_{\rm IR}}{\xi^2+\omega_{\rm IR}^2}+
\frac{\omega_{\rm UV}^2C_{\rm UV}}{\xi^2+\omega_{\rm UV}^2},
\label{eq36}
\end{equation}
\noindent
where $C_{\rm UV} = 5.07$, $\omega_{\rm UV} = 0.735 \times 10^{16}$~rad/s,
$C_{\rm IR} = 108$,
and $\omega_{\rm IR} = 0.7 \times 10^{14}$~rad/s. The resulting static permittivity
of TiO$_2$ is $\varepsilon_0^{(-1)} = \varepsilon^{(-1)}(0) = 114.07$.
In Fig.~\ref{fg2}, the permittivity of TiO$_2$ as a function of $\xi$ is shown
by the line labeled TiO$_2$.

\begin{figure}[t]
\vspace*{-3.2cm}
\hspace*{-1.9cm}
\includegraphics[width=5.5in]{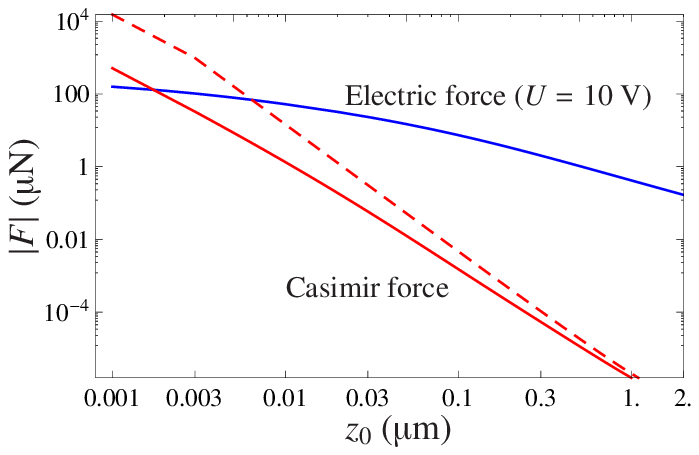}
\vspace*{-12.cm}
\caption{\label{fg3}
The magnitudes of the Casimir and electric forces acting between the
smooth surfaces of a membrane and a bottom electrode of the MEMS capacitive
switch are shown as the functions of separation by the bottom and top solid
lines, respectively. The Casimir force between the membrane and the bottom
electrode made of ideal metal is shown by the dashed line.
}
\end{figure}

We start from calculating the Casimir force acting between the smooth surfaces
of a membrane and a bottom electrode. The computations at $z_0 \geq 3$~nm are
performed by Eq.~(\ref{eq19}) at room temperature $T = 300$~K using the
reflection coefficients defined in Eqs.~(\ref{eq3}) -- (\ref{eq5}) with the
values of the switch parameters indicated above and the dielectric
permittivities of Au, Al, and TiO$_2$ along the imaginary frequency axis
shown in Fig.~\ref{fg2}. At $1~{\rm nm} \leq z_0 \leq 3$~nm computations
are performed using Eq.~(\ref{eq25}). The computational results for $F_C$
as a function of the closest separation $z_0$ are shown by the bottom
solid line in Fig.~\ref{fg3}.

As is seen in Fig.~\ref{fg3}, at $z_0 = 1~\muup$m the Casimir force is
equal to $F_C = -1.57 \times 10^{-6}~\muup$N, i.e., is rather weak, but its
magnitude quickly increases with decreasing separation. Thus, at $z_0 = 100$~nm
and 3~nm, the Casimir force is equal $-1.518 \times 10^{-3}~\muup$N and
$-32.665~\muup$N, respectively.

In many papers, the role of the Casimir force in MEMS switches was determined
considering the boundary surfaces made of ideal metal.\cite{13,14,17,21,23,24,26,27,32}
In recent paper,\cite{89} the Casimir force in arc-plate geometry for Au and
Al membranes in MEMS applications was also calculated using the idealization of
ideal metal. However, at short separations characteristic for the MEMS switches
this idealization is not applicable. As an illustration, we computed the Casimir
force acting in our geometrical configuration between the ideally metallic
surfaces of a membrane and a bottom electrode. For this purpose, one puts in
Eq.~(\ref{eq19})
\begin{equation}
R_{\rm TM}^{(+)}(i\zeta_l,v)R_{\rm TM}^{(-)}(i\zeta_l,v)=
R_{\rm TE}^{(+)}(i\zeta_l,v)R_{\rm TE}^{(-)}(i\zeta_l,v)=1.
\label{eq37}
\end{equation}

The computational results for the magnitude of the Casimir force between an
ideally metallic membrane and a bottom electrode are shown in Fig.~\ref{fg3}
by the dashed line as a function of $z_0$. It is seen that at $z_0 = 1~\muup$m
the Casimir force between the ideally metallic surfaces is equal to
$F_C^{\rm IM} = -2.06 \times 10^{-6}~\muup$N, i.e., its magnitude is by the factor of
1.31 larger than for real materials. At shorter separations of $z_0 = 100$~nm
and 3~nm, the Casimir force between ideal metal surfaces is equal to
$-4.685 \times 10^{-3}~\muup$N and $-977.21~\muup$N, respectively. The magnitudes of
these forces exceed the above results obtained for real materials by the factors
of 3.1 and 29.9, respectively. Thus, in MEMS applications, it is not possible
to use the idealization of ideal metals when calculating the Casimir force
(in Ref.~\onlinecite{89} this idealization is used at separations down to 100~nm).

We are coming now to calculation of the electric force acting between the switch
surfaces of a membrane and a bottom electrode. The computations were performed
by Eq.~(\ref{eq31}) with the same parameters of the capacitive MEMS switch as
used above and,
\tm{as an example,}
the value of the applied voltage $U = 10$~V. The
computational results are shown in Fig.~\ref{fg3} by the top solid line as the
function of separation.

As is seen in Fig.~\ref{fg3}, at all separations exceeding 1.9~nm the magnitudes
of the electric force far exceed those of the Casimir force. Thus, at $1~\muup$m
and 3~nm the electric force is equal to $-0.21~\muup$N and $-102.76~\muup$N,
respectively. These are larger in magnitude than the values of the Casimir force
by the factors of $1.34 \times 10^5$ and 3.14, respectively. However, at $z_0 = 1.9$~nm
the Casimir force becomes equal to the electric one and exceeds its magnitude at
$z_0 < 1.9$~nm. As an example, at $z_0 = 1$~nm it holds $F_C = -509.2~\muup$N
and $F_{\rm el} = -158.2~\muup$N. Thus,
\tm{with the applied voltage of $U = 10 ~$V, }
the magnitude of the Casimir force is by a
factor of 3.22 larger than the magnitude of the electric one. This should be
taken into account when placing the requirements upon the elastic properties of
a membrane which provide the stable cyclic functioning of the MEMS switch.
\tm{It should be taken into account, however, that, according to Eq. (31),
$F_{\rm el}$ increases as $U^2$ with increasing voltage. Because of this, at
larger actuation voltages up to 50~V used in the capacitive MEMS switches,
the magnitude of the electric force remains greater than the magnitude of the
Casimir force at separations down to 1~nm, i.e., in the region where the RF
signal flow terminates. For instance, with $U = 40~$V, one has
$F_{\rm el} = - 2531.2~\muup$N.}

Next, we consider the impact of surface roughness on the values of the Casimir
and electric forces between a membrane and a bottom electrode. To do this, the
computations of both forces are performed by Eq.~(\ref{eq33}) using the values
of $F_C$ and $F_{\rm el}$ calculated for the smooth surfaces and the models of
roughness considered in Sec.~IV. First, we consider the model 1 of roughness
on the Au surface of a membrane and the single model of roughness on a TiO$_2$
plate (see the specific values of $w_i^{(n)}$ and $H_i^{(n)}$ for these models
presented in Sec.~IV).

\begin{figure}[b]
\vspace*{-4cm}
\hspace*{-1.9cm}
\includegraphics[width=5.5in]{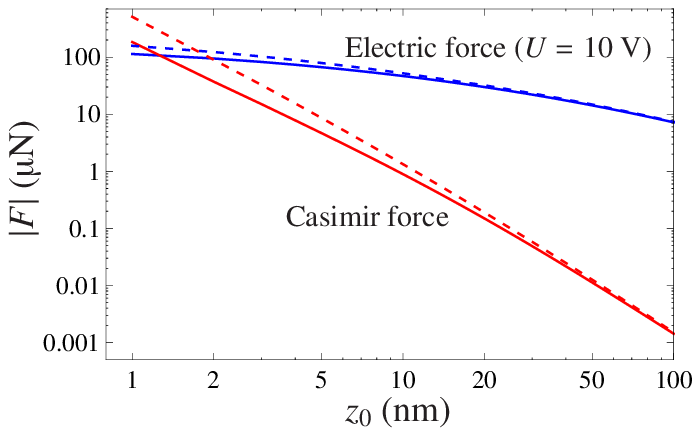}
\vspace*{-12.cm}
\caption{\label{fg4}
The magnitudes of the Casimir and electric forces acting between the
rough surfaces of a membrane (roughness model 1) and a bottom electrode of
the MEMS capacitive switch are shown as the functions of separation by the
bottom and top solid lines, respectively. The bottom and top dashed lines
show the same forces acting between smooth surfaces.
}
\end{figure}
In Fig.~\ref{fg4}, the Casimir and electric forces computed between the rough
surfaces of a membrane (model 1) and a bottom electrode are shown by the
bottom and top solid lines, respectively, as the functions of separation.
The bottom and top dashed lines reproduce the bottom and top solid lines
from Fig.~\ref{fg3}, which show the Casimir and electric forces, respectively,
computed for the smooth surfaces.

As is seen in Fig.~\ref{fg4}, the account of surface roughness decreases the
magnitudes of the Casimir and electric forces. Thus, at $z_0 = 3$~nm,
$F_C^R = -15.5~\muup$N and $F_C = -32.66~\muup$N, i.e., the force magnitude
computed with account of roughness is by the factor of 2.11 less. At $z_0 = 1$~nm,
we have $F_C^R = -184.1~\muup$N and $F_C = -509.2~\muup$N. Thus, for rough surfaces,
the force magnitude is by the factor of 2.76 smaller.

In a similar way, for the electric force one obtains $F_{\rm el}^R = -82.54~\muup$N
and $F_{\rm el} = -102.76~\muup$N at $z_0 = 3$~nm and $F_{\rm el}^R = -111.72~\muup$N and
$F_{\rm el} = -158.2~\muup$N at $z_0 = 1$~nm. Thus, for rough surfaces at $z_0 = 1$ and
3~nm, the force magnitude is by the factors of 1.24 and 1.42 smaller, respectively.
\tm{For larger operating voltages, one obtains similar results. For instance,
for $U = 40~$V at $z_0 =1~$nm, $F_{\rm el}^R = - 1787.52~\muup$N to compare
with $F_{\rm el} = - 2531.2~\muup$N, i.e., by again the factor of 1.42 decreased
force magnitude.}
It is seen, however, that the impact of surface roughness
on the electric force is somewhat smaller than on the Casimir one.

From Fig.~\ref{fg4} it is also seen that the impact of surface roughness on both
the Casimir and electric forces decreases with increasing separation. Thus, at
$z_0 = 100$~nm, $F_C^R = -1.43 \times 10^{-3}~\muup$N and
$F_C = -1.52 \times 10^{-3}~\muup$N resulting in by the factor of 1.06 smaller
force magnitude due to the impact of roughness. In a similar way, for the
electric force, at the same separation it holds $F_{\rm el}^R = -7.34~\muup$N and
$F_{\rm el} = -7.36~\muup$N, i.e., by only the factor of 1.003 smaller force
magnitude between the rough surfaces.

According to Fig.~\ref{fg4}, for all $z_0 > 1.55$~nm, $|F_{\rm el}^R > |F_C^R|$ and
for $z_0 < 1.55$~nm the magnitude of the Casimir force exceeds the magnitude
of the electric one, $|F_C^R > |F_{\rm el}^R|$. Thus, under the impact of surface
roughness, the value of $z_0$, which equates the magnitudes of both forces
becomes less. At $z_0 = 1$~nm one has $F_C^R = -184.1~\muup$N and
$F_{\rm el}^R = -111.72~\muup$N, i.e., the magnitude of the Casimir force exceeds
the magnitude of the electric one by the factor of 1.65.

\begin{figure}[t]
\vspace*{-1cm}
\hspace*{-1.9cm}
\includegraphics[width=5.5in]{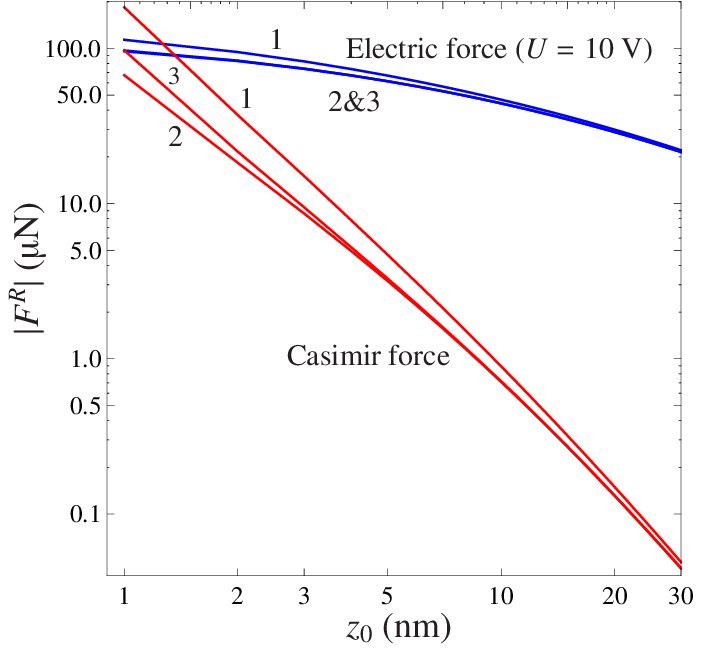}
\vspace*{-12.cm}
\caption{\label{fg5}
The magnitudes of the Casimir and electric forces acting between the
rough surfaces of a membrane (roughness models 2, 3, and 1) and a bottom
electrode of the MEMS capacitive switch are shown as the functions of
separation by the three bottom and two top solid lines, respectively (see
the text for further discussion).
}
\end{figure}

Below we demonstrate that the use of different models of roughness on the Au
surface of a membrane leads to only quantitative changes in the above results.
In Fig.~\ref{fg5}, the computational results for the Casimir and electric
forces obtained by using different models of roughness are shown as the
functions of separation. The top lines for both forces reproduce the solid
lines plotted in Fig.~\ref{fg4}, which were obtained using the roughness
model 1. The bottom and middle lines in Fig.~\ref{fg5} are obtained using the
models of roughness on Au surface 2 and 3, respectively. In all these cases,
the unique model of roughness on the TiO$_2$ surface described in Sec. IV
has been used. Note also that for the electric force the lines obtained using
the roughness models 2 and 3 overlap.

As is seen in Fig.~\ref{fg5}, the major decrease in the magnitude of the
Casimir force occurs when the roughness model 2 is used (the bottom line in
Fig.~\ref{fg5}). Thus, at $z_0 = 3$~nm and 1~nm $F_C^R = -8.64~\muup$N and
$-67.21~\muup$N, respectively. Comparing these values with the corresponding
above values for smooth surfaces, we find that for this model of roughness
the magnitude of the Casimir force decreases by the factors of 3.78 and 7.58,
respectively.

The magnitudes of the electric force computed using the roughness models 2
and 3 are also smaller than those computed using the model 1. At $z_0 = 3$~nm,
$F_{\rm el}^R = -74.14~\muup$N and $-73.94~\muup$N for the roughness models 2
and 3, respectively. This should be compared with the value of
$F_{\rm el} = -102.76~\muup$N for the smooth surfaces leading to by the factors
of 1.386 and 1.390 decreased force magnitudes for the models 2 and 3,
respectively. At $z_0 = 1$~nm, for the models 2 and 3, it holds
$F_{\rm el}^R = -95.88~\muup$N and $-96.98~\muup$N, respectively. Comparing
these values with $F_{\rm el} = -158.20~\muup$N for smooth surfaces, one arrives
to by the factors of 1.65 and 1.63 decreased force magnitudes due to the
impact of surface roughness. As is seen in Fig.~\ref{fg5}, for the roughness
model 3 the magnitude of the Casimir force reaches the magnitude of the
electric one at $z_0=1~$nm. However, for the roughness model 2, the magnitude
of the Casimir force remains smaller than the magnitude of the electric force
at all $z_0 \geq 1$~nm.

With increasing separation, the differences in the force magnitudes computed
using different models of surface roughness disappear. Thus, at $z_0 = 30$~nm
the electric force computed using the model 1 is equal to
$F_{\rm el}^R = -22.08~\muup$N and by the models 2 and 3 is very close to this
value, $F_{\rm el}^R = -21.52~\muup$N. The Casimir force computed using the
model 1 at $z_0 = 30$~nm is $F_{C}^R = -4.9 \times 10^{-2}~\muup$N and
using the models 2 and 3 is $F_{C}^R = -4.47 \times 10^{-2}~\muup$N.

\tm{The above results were obtained for the operating voltage of $U = 10~$V.
As already mentioned, for the relatively larger operating voltages the magnitude
of the electric force remains larger than the magnitude of the Casimir one at
separations down to 1 nm. However, the qualitative results concerning an
impact of surface roughness on the electric force remain unchanged.}

\section{CONCLUSIONS AND DISCUSSION}

In the foregoing, we have calculated both the Casimir (van der Waals) and
electric forces acting in the MEMS capacitive switches with a cylindrical
membrane taking into account the geometry of a membrane, the real properties
of both the membrane and bottom electrode materials, and the surface roughness.
This calculation was performed using the Lifshitz theory in its application
area, the proximity force approximation, and, at the shortest separations
between a membrane and a bottom electrode, the
\tm{phenomenological}
method of additive summation
of interatomic van der Waals potentials corrected for the effects of
nonadditivity.

It was shown that, at separations of about 1~nm, where the membrane is
effectively in contact with the transmission line (high capacitance) and
the RF signal is terminated, the magnitude of the Casimir force can reach
the magnitude of the electric one and may even exceed it depending on the
character of surface roughness
\tm{and the value of the operating voltage.}
This means that the Casimir force must be
taken into account when selecting the material and thickness of a membrane.
In fact the restoring elastic force, which brings the membrane in its initial
position after switching off the operating voltage, should be larger than
the magnitude of the attractive Casimir force. In other case, under the
impact of the latter, membrane would collapse on the bottom electrode and
the capacitive switch will become and remain nonoperative.

According to our results, for the typical parameters of the MEMS capacitive
switch used in computations, the magnitude of the Casimir force at the
shortest separation of 1~nm between the rough surfaces of a membrane and
a bottom electrode varies between $-67.21~\muup$N and $-184.1~\muup$N
depending on the model of roughness used in computations.
\tm{An account of the actual material properties is even more important because,
already at the separation of 3 nm, the Casimir forces computed using the real
material properties of the switch elements and the idealization of ideal metal
differ by the factor of 29.9. Note that an inaccuracy of the approximate method
used at separations below 1 nm leads to only small errors in the value of
numerical coefficient in the force dependence on separation, and this results
in much smaller changes in the Casimir force magnitude than those due to
failure to take account of the realistic material properties and surface roughness.}

In practical
applications, one should, first, to investigate the character of surface
roughness by means of an atomic force microscope and, then, calculate the
Casimir (van der Waals)  force taking the surface roughness into account
by means of the formalism presented in this paper. Next, the value of the
necessary thickness of a membrane should be calculated in order the
restoring elastic force be greater than the magnitude of the Casimir force
in the contact position of a membrane with a bottom electrode. This can be
done using the much developed formalism of Euler-Bernoulli beam theory
applied to the investigation of MEMS switches in many
papers.\cite{13,14,17,21,23,24,26,27,32,89}
\tm{The pull-in phenomena in the capacitive switches are often attributed to the role of electric charges, which may be localized on a dielectric plate covering the central signal line. We show that another cause leading to the pull-in phenomena in capacitive switches is the unavoidable Casimir force if the restoring elastic force does not exceed its magnitude at the closest separation between the membrane and the bottom electrode.}

To conclude, we have proven that the Casimir force in MEMS switches should
be calculated using the realistic material properties of the boundary
materials because the idealization of ideal metal employed in many papers
on the subject\cite{13,14,17,21,23,24,26,27,32,89} leads to the values of
the Casimir force significantly different from the correct ones. The same
is true regarding an account of surface roughness, which decreases the
magnitude of the Casimir force for a given separation between the closest
points on the interacting surfaces. Thus, both the real material properties
and surface roughness should be taken into account for determining the
correct value of the Casimir force and, finally, the necessary thickness
of the membrane and value of the elastic force restoring it in the initial
position after switching off the operating voltage.

By and large, the presented formalism may find applications when developing
the MEMS capacitive switches of next generations stable against the pull-in
phenomena.

\begin{acknowledgments}
This work was supported by the State Assignment for Basic Research
(project FSEG-2026-0018).
\end{acknowledgments}

\end{document}